\documentclass[10pt,prl,twocolumn,floatfix,tightenlines,showpacs,showkeys,preprintnumbers,altaffilletter, nofootinbib, superscriptaddress, longbibliography,bibnotes]{revtex4-1}

\pdfoutput=1
\usepackage[colorlinks=true,breaklinks=true]{hyperref}
\hypersetup{allcolors=[rgb]{0.0 0.0 0.6},linkcolor=[rgb]{0.8 0.0 0.5}}
\usepackage{amsmath,amssymb}
\usepackage{epsfig}  
\usepackage{graphicx}   
\usepackage{commath}
\usepackage{slashed}             
\usepackage{url}
\usepackage{color}
\usepackage{multirow}
\usepackage{placeins}
\usepackage[dvipsnames]{xcolor}
\usepackage{soul}
\usepackage[normalem]{ulem}

\allowdisplaybreaks

\bibpunct{[}{]}{,}{n}{}{,}

\newcommand\barparenb[1]{\overset{%
   \scalebox{0.4}{$(\mkern-1mu-\mkern-1mu)$}}{#1}}

\begin{document}

\title{Collisional triggering of fast flavor conversions of supernova neutrinos}
             
\author{Francesco Capozzi}
\email{capozzi@mppmu.mpg.de}
\affiliation{Max Planck Institut f\"{u}r Physik (Werner-Heisenberg-Institut), F\"{o}hringer Ring 6, 80805 M\"{u}nchen, Germany. 
}           
\author{Basudeb Dasgupta}
\email{bdasgupta@theory.tifr.res.in}
\affiliation{Tata Institute of Fundamental Research,
             Homi Bhabha Road, Mumbai 400005, India.}
\author{Alessandro Mirizzi}
\email{alessandro.mirizzi@ba.infn.it }
\affiliation{Dipartimento Interateneo di Fisica ``Michelangelo Merlin'', Via Amendola 173, 70126 Bari, Italy.}
\affiliation{Istituto Nazionale di Fisica Nucleare - Sezione di Bari,
Via Amendola 173, 70126 Bari, Italy.}
\author{Manibrata Sen}
\email{manibrata.sen@gmail.com }  
\affiliation{Tata Institute of Fundamental Research,
             Homi Bhabha Road, Mumbai 400005, India.}
\author{G{\"u}nter Sigl}
\email{guenter.sigl@desy.de}  
\affiliation{II. Institute for Theoretical Physics, Hamburg University
Luruper Chaussee 149, D-22761 Hamburg, Germany}

\preprint{MPP-2018-206, TIFR/TH/18-27}

\date{\today}

\begin{abstract}
Fast flavor conversions of supernova neutrinos, possible near the neutrinosphere, depends on an interesting interplay of collisions and neutrino oscillations. Contrary to naive expectations, the rate of self-induced neutrino oscillations, due to neutrino-neutrino forward scattering, comfortably exceeds the rate of collisions even deep inside the supernova core. Consistently accounting for collisions and oscillations, we present the first calculations to show that collisions can create the conditions for fast flavor conversions of neutrinos, but oscillations can continue without significant damping thereafter. This may have interesting consequences for supernova explosions and the nature of its associated neutrino emission.
\end{abstract}

\maketitle

 \emph{Introduction.\,}--  Neutrinos emitted from supernovae can undergo significant changes in their flavor composition due to the large densities inside the star. This enhanced flavor conversion is caused by the refractive potentials, due to forward scattering of the $\nu$ (or $\bar{\nu}$) off the $e^-$ and the other $\nu$ and $\bar{\nu}$ in the background, while non-forward collisions typically damp the flavor oscillations. In particular, self-induced or collective oscillations associated with neutrino-neutrino forward-scattering have been a source of many interesting and puzzling effects~\mbox{\cite{Duan:2006an,Hannestad:2006nj,Fogli:2007bk,Dasgupta:2009mg}}. See refs.~\mbox{\cite{Duan:2010bg,Mirizzi:2015eza,Chakraborty:2016yeg}} for recent reviews. The most puzzling manifestation of these collective effects was pointed out by Raymond Sawyer~\cite{Sawyer:2005jk,Sawyer:2008zs}, who argued that the growth rate of flavor conversions can be proportional to the neutrino potential $\mu \sim \sqrt{2} G_F n_\nu$, which scales with the neutrino density $n_\nu$ but is little influenced by neutrino masses after the onset of flavor conversions.
 Thus, this \emph{fast} oscillation rate can greatly exceed the ordinary neutrino oscillation frequency $\omega=\Delta m^2/(2E)$, by a factor of $\mu/\omega\sim10^5$, and occur in regions much deeper in the star, from where neutrinos are emitted.

Fast neutrino conversions can take place only if the electron lepton number (ELN) distribution, i.e., the difference of the $\nu_e$ and $\bar{\nu}_e$ angular fluxes, changes its sign across some direction of emission at any given point inside the supernova~\cite{Sawyer:2015dsa, Chakraborty:2016lct,Dasgupta:2016dbv,Izaguirre:2016gsx,Capozzi:2017gqd,Dasgupta:2017oko, Abbar:2017pkh}. This necessary condition for fast oscillations, that the ELN has a ``crossing'' through zero, obviously requires that $\nu_e$ and $\bar{\nu}_e$ have different collision rates. Such a difference is quite likely near the neutrino decoupling region in a supernova: due to neutron richness of stellar matter the ${\bar\nu}_e$ decouples earlier, so the ${\bar\nu}_e$ are more forward peaked than $\nu_e$\,; further, if number density of $\nu_e$ does not greatly exceed that of $\bar{\nu}_e$, i.e., the lepton asymmetry is modest, the ELN could exhibit the required crossing. However, this also immediately raises a red flag -- if collisions are important to create the conditions for fast conversions, wouldn't they damp oscillations too?

In this \emph{Letter}, we present the first calculations to explain the interplay of collisions and fast oscillations. We note that the collision rates $\Gamma_{\nu_e}$ and $\Gamma_{\bar{\nu}_e}$ are significantly smaller that the refractive potential $\mu$ even inside the supernova core. As a result, collisions are dominant only initially and can create conditions for fast oscillations when oscillations are not yet operative. However, once fast oscillations have been triggered, the collision rates being smaller than $\mu$, are not large enough to lead to damping of oscillations. If these fast oscillations occur in supernovae~\cite{Sawyer:2005jk,Sawyer:2008zs, Sawyer:2015dsa, Chakraborty:2016lct,Dasgupta:2016dbv,Izaguirre:2016gsx,Capozzi:2017gqd,Dasgupta:2017oko, Abbar:2017pkh}, they represent a major change to the existing paradigm wherein the collisional and free-streaming regimes are believed to be well-separated. Our results suggest that this simplification may not always hold, with potentially important consequences for supernova astrophysics and neutrino physics.

\begin{figure}[!b]
\begin{centering}
\includegraphics[width=0.6\columnwidth]{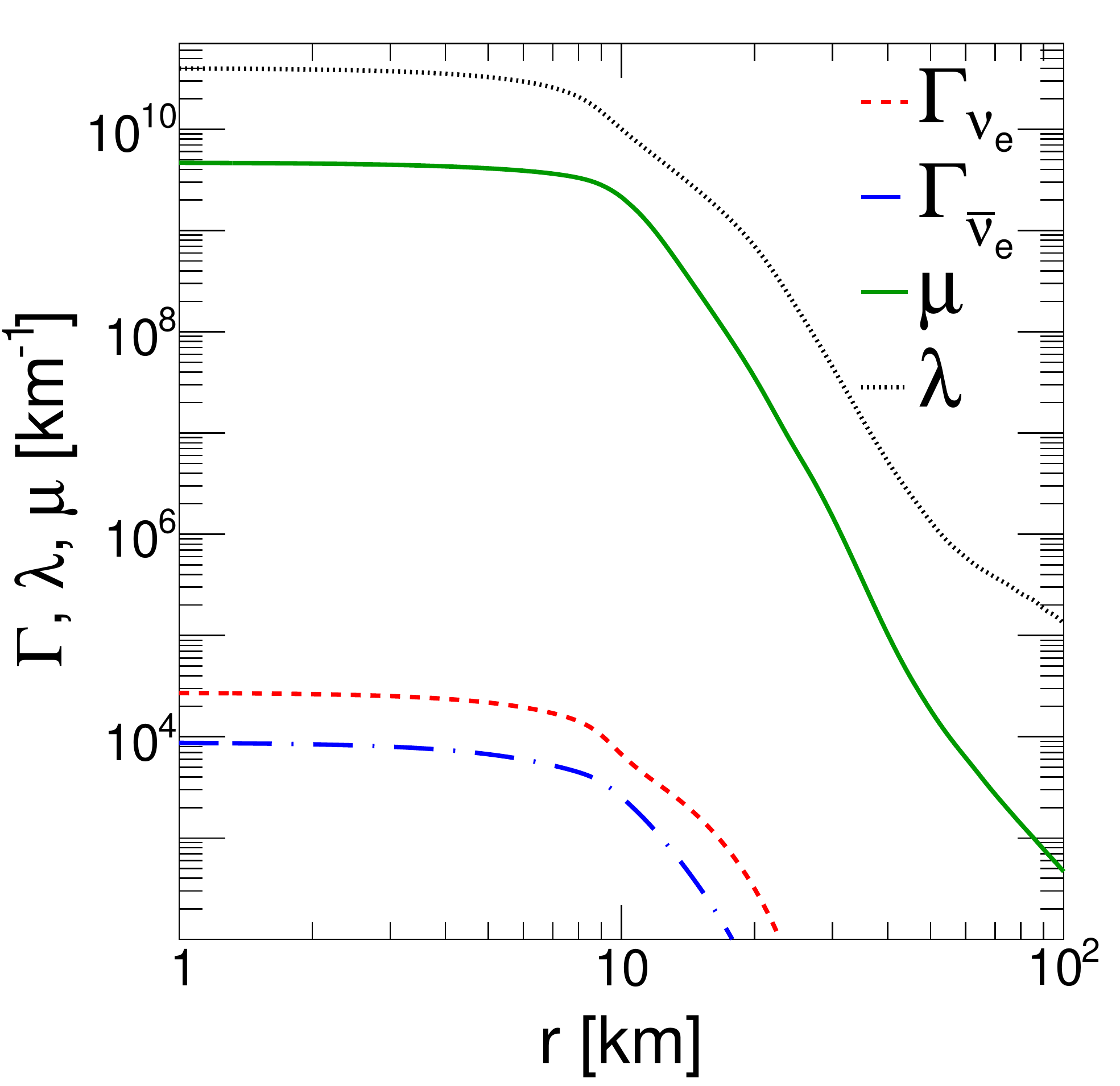}
\end{centering}
\caption{Properties of an 11 $M_{\odot}$ supernova derived from the simulation in ref.~\cite{Tamborra:2017ubu}. Radial profiles of neutrino potential $\mu$, matter potential $\lambda$, and scattering rates $\Gamma_{\nu_e}$ and $\Gamma_{\bar\nu_e}$, are shown
for a snapshot at a post-bounce time $t=0.170$~s.}
\label{fig:profiles}
\end{figure}  

We will start with a simple illustration of the relevant scales in the problem. Fig.\,\ref{fig:profiles} shows the neutrino potential $\mu$, and the charged-current collisional rates $\Gamma \sim n_B \sigma$, where $n_B$ is the nucleon density and $\sigma$ is the charged-current cross section for the $\nu_e$ and $\bar\nu_e$, from an $11\,M_{\odot}$ spherically symmetric (1D) supernova model, simulated by the Garching group, at a post-bounce time $=0.170$\,s~\cite{Tamborra:2017ubu}. These quantities are energy-averaged; see the Supplemental Material for details. As apparent, the neutrino potential $\mu$ is always larger than the $\nu_e$ collisional rate, by no less than $\sim 4$ orders of magnitude. The $\bar\nu_e$ collisional rate is $\sim 3$ times smaller than the $\nu_e$ collision rate. Thus, even in the deepest regions, at $r \lesssim$ 10 km where these quantities become roughly constant, the refractive effects remain stronger. The matter term $\lambda = \sqrt{2} G_F n_e$ is one order of magnitude larger than $\mu$, but it is now understood that flavor conversions can grow locally and are not suppressed by a large matter effect~\cite{Dasgupta:2016dbv, Abbar:2017pkh}, unless stationarity is imposed by fiat~\cite{Airen:2018nvp}. Therefore, if fast conversions are triggered somewhere in the neutrino decoupling region, they may affect the entire region near the neutrinosphere. However, this physics has not been explored. Supernova simulations assume that oscillations do not take place deep in the star, while oscillation calculations completely ignore collisions even when considering fast conversions. In this study, we relax these assumptions and demonstrate the interplay of collisions and oscillations in a toy-model. In the following, we set up the equations, define our toy-model and present the numerical results for the same, and conclude by discussing their relevance to supernova physics.

\emph{Equations of motion including collisions.\,}-- Ignoring external forces, the equations of motion (EoMs) for the $\nu$ occupation number matrices $\varrho_{{\bf p}, {\bf x},t}$ for momentum ${\bf p}$ at position ${\bf x}$ and time $t$ are~\cite{Sigl:1992fn,McKellar:1992ja,Vlasenko:2013fja,Blaschke:2016xxt,Volpe:2013jgr} 
\begin{equation}
\left(\partial_t + {\bf v}_{\bf p} \cdot \nabla_{\bf x}\right) \varrho_{{\bf p}, {\bf x},t} 
= - i [\Omega_{{\bf p}, {\bf x},t}, \varrho_{{\bf p}, {\bf x},t}] + {\cal C} [\varrho_{{\bf p}, {\bf x},t}]
\,\ ,
\label{eq:eom}
\end{equation}
where, in the Liouville operator on the left-hand side,  the first term accounts for explicit time dependence, while the second term, proportional to the neutrino velocity ${\bf v_p}$, encodes the spatial dependence due to particle free streaming. 
 The right-hand-side contains the oscillation Hamiltonian $\Omega_{{\bf p}, {\bf x},t}$, which is a sum of the vacuum term depending on the mass-squared matrix of neutrinos, the matter term depending on background density of electrons, and the self-interaction term\,\mbox{$\int d^3{\bf q}/(2\pi)^3 (1-{\bf v}_{\bf p}\cdot {\bf v}_{\bf q})({\varrho}_{{\bf q}, {\bf x},t}-{\bar\varrho}_{{\bf q}, {\bf x},t})$}~\cite{Mirizzi:2015eza}. The last term on the right-hand-side of Eq.\,(\ref{eq:eom}) accounts for non-forward collisions. Antineutrinos represented by ${\bar\varrho}_{{\bf p}, {\bf x},t}$ obey the same equation, but with an opposite sign for the vacuum oscillation term. 

Our goal here is to capture the main features of the interplay between flavor conversions and collisions. Therefore, we simplify the collisional term as described below. We include only the charged-current absorption and emission processes that create $\nu_e$ and $\bar{\nu}_e$ and their flavor and angular asymmetries, neglecting neutral-current interactions that both produce the other flavors and affect kinetic decoupling~\cite{Raffelt:2001kv, Keil:2002in}. The relevant collisional term, derived in~\cite{Raffelt:1992bs}, can be mimicked by~\cite{Dolgov:2001su}
\begin{equation}
{\cal C} [\varrho_{{\bf p}, {\bf x},t}]
=\frac{1}{2}\left\{\varGamma_{{\bf p}}, (\varrho^{{\rm eq}}_{\bf p}-\varrho_{\bf p}) \right\}  \,\ , 
\label{eq:coll}
\end{equation}
where $\{,\}$ denotes an anticommutator and $\varrho^{{\rm eq}}_{\bf p}$ represents the equilibrium value of the occupation matrix  and takes into account Pauli-blocking effects. The matrix $\varGamma={\rm diag}(\Gamma_{\nu_e},0)$ in the flavor basis has a nonzero contribution only for the electron flavor and is proportional to the collision rate $\Gamma_{\nu_e}$ for the processes allowed, e.g., $p\,e^- \to n\,\nu_e$ for $\nu_e$. Analogously, only the process $n\,e^+ \to p\,\bar{\nu}_e$ is relevant for $\bar{\nu}_e$. These rates in a supernova model are shown in Fig.\,\ref{fig:profiles}.

The collisional term in Eq.\,(\ref{eq:coll}) is analogous to the one used in the context of neutrino flavor conversions in the early Universe~\cite{Dolgov:2002ab}. It has a two-pronged effect. It populates the diagonal components of $\varrho_{\bf p}$; in particular, if $\varrho^{\rm eq}_{\bf p}$ is not the same for all modes ${\bf p}$ these states get differently populated. However, it damps the off-diagonal terms of of $\varrho_{\bf p}$, destroys coherence, and inhibits any kind of flavor oscillation if sufficiently strong.

\begin{figure}[!b]
\begin{centering}
\includegraphics[width=0.95\columnwidth]{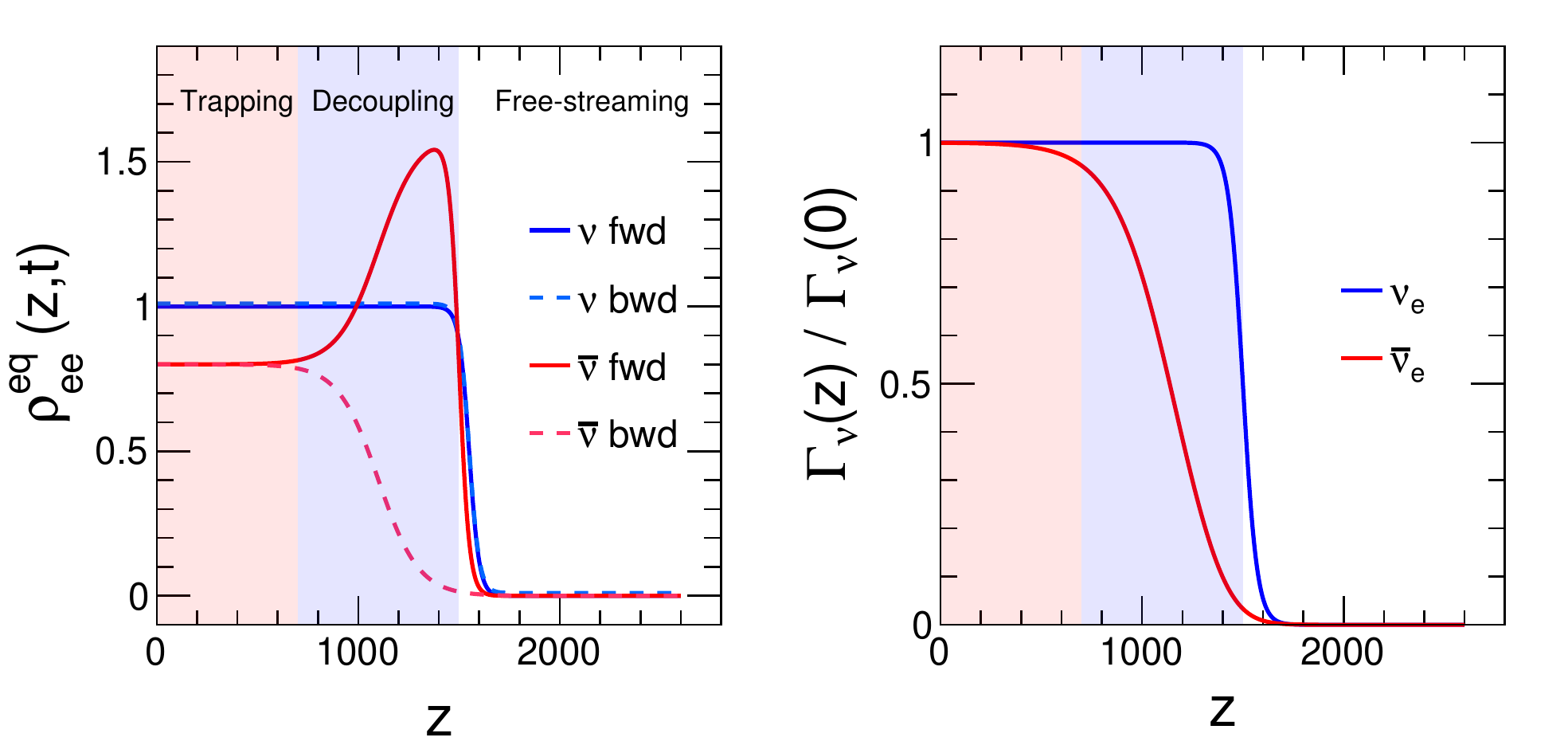}
\end{centering}
\caption{Equilibrium abundances and relative collision rates for $\nu_e$ and $\bar{\nu}_e$. Left panel: Equilibrium values of the occupation numbers, $\varrho^{\rm eq}_{ee}$, for $\nu_e$ and $\bar\nu_e$ in the forward and backward directions. Right panel: Collision rates $\Gamma_{\nu_e}$ and $\Gamma_{\bar\nu_e}$. }
\label{fig:init}
\end{figure}  

\emph{Numerical examples.\,}-- We consider time-dependent flavor evolution in one spatial dimension labeled by $z$, mimicking the temporal and radial flavor evolution in a spherically symmetric supernova. Further, we take only two momentum modes of equal energy, counter-propagating in the forward ($p_z>0$) and backward ($p_z<0$) directions, labelled by $f$ and $b$, respectively. Their equilibrium abundance profiles, without oscillations, $\barparenb\varrho{\raisebox{3pt}{\scriptsize{\rm \,eq}}}{\raisebox{-3pt}{\hspace{-0.3cm}\tiny$f$\hspace{0.2cm}}}$ and $\barparenb\varrho{\raisebox{3pt}{\scriptsize{\rm \,eq}}}{\raisebox{-3pt}{\hspace{-0.3cm}\tiny$b$\hspace{0.2cm}}}$, are enforced to have a crossing in the ELN; equivalent to assuming different decoupling profiles. We then numerically solve the non-linear EoMs [Eq.~(\ref{eq:eom})] in $z\in [0,\,L]$ and $t$.
To emphasize the natural scale of the problem, we express all quantities in units of a scale $\mu_0$. Given the huge dynamic range between $\mu$ and $\Gamma_{\nu}$, one cannot simulate the SN model of Fig.\,\ref{fig:profiles} in complete detail. We assume quasi-instantaneous decoupling, and model the decoupling region as a box with $L=2800$, which e.g. for $\mu_0=10^{5}$~km$^{-1}$ is $\mathcal{O}(10^{-2}\,{\rm km})$ in size -- somewhat smaller than in a supernova. In this box, we take $\mu$ to be spatially constant but $\Gamma_{\nu}(z)$ to have the profile shown in Fig.\,\ref{fig:init}, qualitatively encoding the decoupling behavior. Results will be largely independent on neutrino mass-mixing parameters, which only affect the seed for the flavor conversions. Nonetheless, for concreteness, we use an inverted mass ordering, with the vacuum oscillation frequency $\omega= 5\times10^{-5}$, which for $\mu_0=10^{5}$~km$^{-1}$ corresponds to the atmospheric neutrino mass-square difference $\Delta m^2=2.4\times 10^{-3}\,{\rm eV}^2$ with a representative neutrino energy $E=12$\,MeV. We set the matter term $\lambda$ to zero, for simplicity, while using a matter-suppressed mixing angle $\theta=10^{-3}$. The neutrino velocities are taken to be $v_f=-v_b=0.2$, a ballpark value affecting only the propagation speed of the flavor instability. The Supplemental Material has details of the numerical methods.

\begin{figure*}[!t]
\begin{centering}
\includegraphics[width=0.85\textwidth]{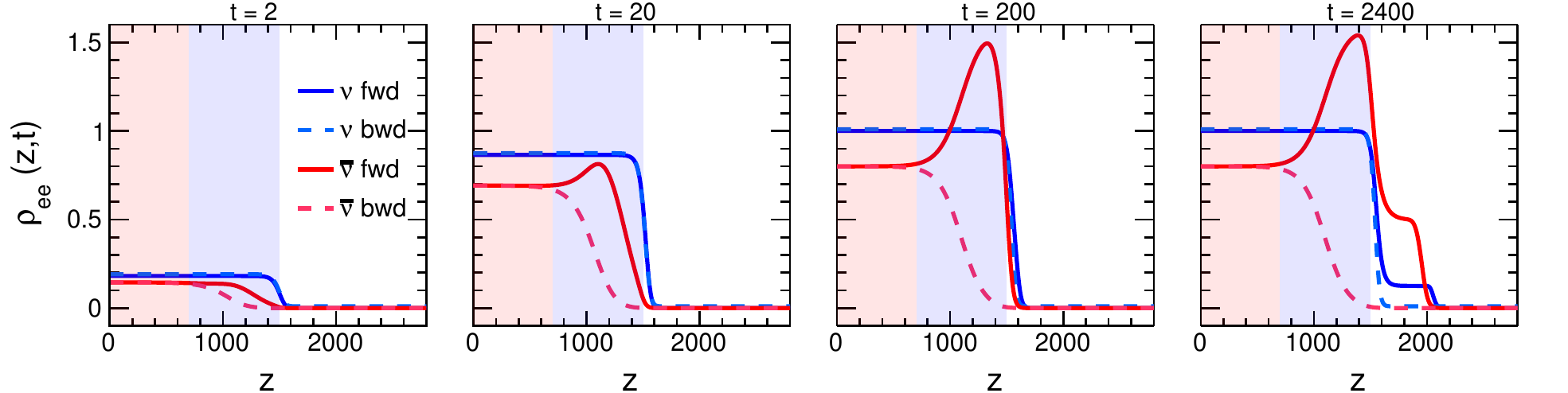}
\end{centering}
\caption{Two beam model in the no-oscillation limit, with $\Gamma_{\nu_e}(0)=\Gamma_{\bar{\nu}_e}(0)=0.1$ and $\mu=0$. Evolution of forward and backward going mode-occupations for $\nu_e$ and $\bar\nu_e$, as a function of $z$ for different representative times. Note the approach to equilibrium, followed by free-streaming in the right-most zone at late times.
} 
\label{fig:xc01mu0}
\end{figure*}
\vspace{0.2cm}  

\begin{figure*}[!t]
\begin{centering}
\includegraphics[width=0.85\textwidth]{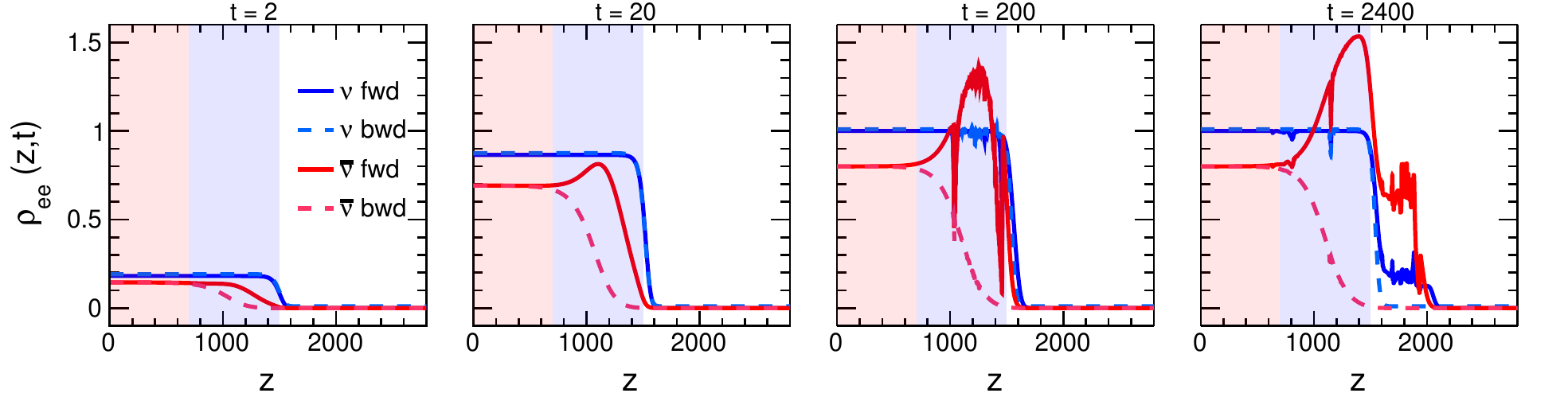}
\end{centering}
\caption{Two beam model with $\Gamma_{\nu_e}(0)=\Gamma_{\bar{\nu}_e}(0)=0.1$ and $\mu=1$. Evolution of forward and backward going mode-occupations for $\nu_e$ and $\bar\nu_e$ as a function of $z$ for different representative times. Note the instability in the $t = 200$ snapshot and that, except for a leading transient, $\rho_{ee}$ approaches approximately the same equilibrium as in Fig.\,\ref{fig:xc01mu0}.}
\label{fig:xc01}
\end{figure*}  
\vspace{0.2cm}

\begin{figure*}[!t]
\begin{centering}
\includegraphics[width=0.85\textwidth]{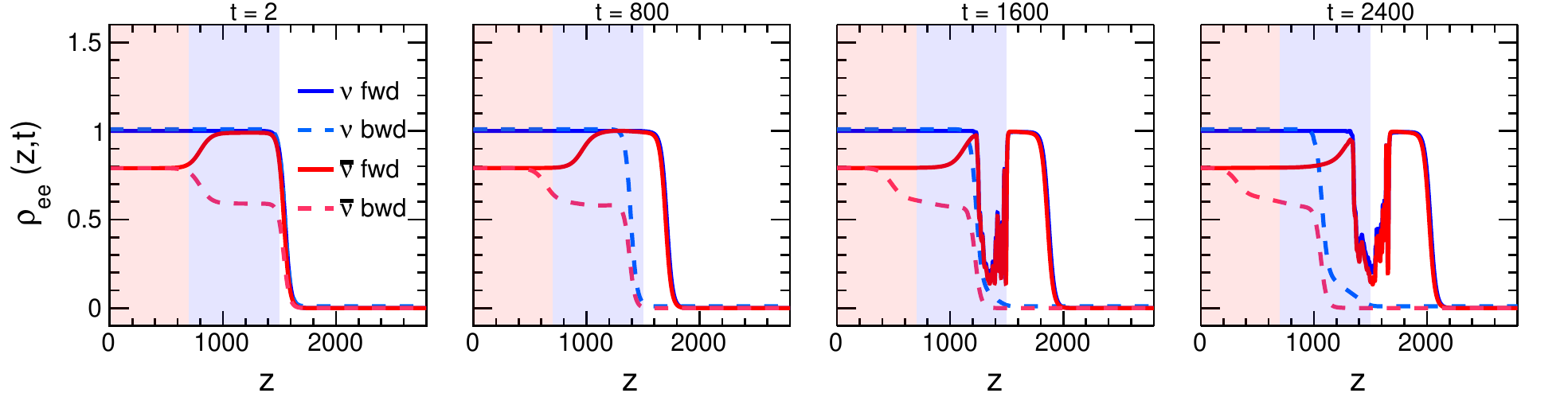}
\end{centering}
\caption{Two beam model with $\Gamma_{\nu_e}(0)=\Gamma_{\bar{\nu}_e}(0)=10^{-4}$ and $\mu=1$. Evolution of forward and backward going mode-occupations for $\nu_e$ and $\bar\nu_e$ as a function of $z$ for different representative times. Note that the instability in the $t=1600$ snapshot is not suppressed even at later times and a different equilibrium is reached after a longer transient.}
\label{fig:xc00001}
\end{figure*}

In the left panel of Fig.\,\ref{fig:init} we plot the equilibrium value of the occupation numbers for $\nu_e$ and $\bar\nu_e$ in the forward and backward directions in the box $z \in [0,2800]$. The box has three zones: $z< 700$ that represents the trapping zone where both $\nu_e$ and $\bar\nu_e$ have equally populated forward and backward modes; $700 < z< 1500$ representing the decoupling zone where $\bar\nu_e$ decouples, while $\nu_e$ decouples around $z\approx 1500$; and $z>1500$ the free-streaming zone where both $\nu_e$ and $\bar\nu_e$ having decoupled can now free-stream. The specific values of $z$ demarcating the zones are chosen ad hoc and do not carry any special significance.
For $z<700$ we assume $\barparenb\varrho{\raisebox{3pt}{\scriptsize{\rm \,eq}}}{\raisebox{-3pt}{\hspace{-0.3cm}\tiny$f$\hspace{0.2cm}}} = \barparenb\varrho{\raisebox{3pt}{\scriptsize{\rm \,eq}}}{\raisebox{-3pt}{\hspace{-0.3cm}\tiny$b$\hspace{0.2cm}}}$, with an excess of $\nu_e$ over ${\bar\nu}_e$.  In the decoupling zone, $700 < z< 1500$, $\nu_e$ have no forward-backward asymmetry, whereas to mimic the decoupling of $\bar\nu_e$ they are assumed to have an excess of forward over backward modes keeping their total number in the first and second zone constant. With such a definition of $\varrho^{\text{eq}}$, collisions will eventually generate a crossing in the ELN, i.e.,  an excess of $\nu_e$ over $\bar\nu_e$ in the backward mode, and vice versa for the forward mode at a fixed location. 
The $\nu_e$ equilibrium occupations are normalized to one, and overall factors absorbed in $\Gamma$ and $\mu$.
Note that an excess of $\bar\nu_e$ over $\nu_e$ can lead to relative occupations larger than one. Finally, in the free-streaming zone at $z>1500$ we choose there are no backward modes.  In the right panel of Fig.\,\ref{fig:init} we show the collision rates $\Gamma_{\bar\nu_e}$ and $\Gamma_{\nu_e}$, both normalized to one at their maximum. $\Gamma_{\nu_e}$ is nearly constant up to about $z=1500$, whereas  $\Gamma_{\bar\nu_e}$starts decreasing around $z=700$. For $z>1500$ both neutrinos and antineutrinos are free-streaming, i.e.,  $\Gamma_{\bar\nu_e}= \Gamma_{\nu_e}=0$.
Note that for both $\varrho^{\rm eq}$ and $\Gamma$ we are considering smooth variations between the three zones, since discontinuities may introduce numerical artifacts in the simulations.

In Fig.\,\ref{fig:xc01mu0}  we plot several time snapshots of the evolution of the occupation numbers, including the collisional term with $\Gamma_{\nu_e}(0)=\Gamma_{\bar{\nu}_e}(0)=0.1$, and setting $\mu=0$ (no fast oscillations). We start with no neutrinos in the box at $t=0$, but they get populated through the collisional term. Already at $t=2$ the population of both forward and backward modes for $\nu_e$ and $\bar\nu_e$ start to grow due to the collisional term.  At $t=200$ all modes, including the forward $\bar\nu_e$, have reached their equilibrium value and in the following time snapshot one simply observes the free propagation of the forward modes into the free-streaming zone where $\Gamma=0$. Note that all modes in the range $z\lesssim 1500$ are frozen to their equilibrium value, as the repopulation is efficient.

In Fig.\,\ref{fig:xc01} we switch on the neutrino-neutrino interaction term, $\mu=1$, keeping  $\Gamma_{\nu_e}(0)=\Gamma_{\bar{\nu}_e}(0)=0.1$. Due to the presence of a crossing in the ELN in the decoupling zone, flavor conversions start to develop (notice the wiggles in the snapshot at $t=200$). However, due to the large collisional term, the system quickly tends to equilibrium and at larger times the evolution is very similar to the case with $\mu=0$. The leading peak seen in the $t=2400$ snapshot is the initial transient. 
Note that for $\Gamma=0.1$, a non-negligible production of non-electron neutrinos occurs due to fast conversions (not-shown). However, with larger values of  $\Gamma \gg \mu$ this population would be significantly suppressed.

Finally, in Fig.\,\ref{fig:xc00001} we significantly lower $\Gamma_{\nu_e}(0)$ and $\Gamma_{\bar{\nu}_e}(0)$ to  $10^{-4}$ in order to represent the realistically expected hierarchy between $\Gamma$ and $\mu$, as shown in Fig.\,\ref{fig:profiles}. As the collisional production rate is significantly slower than in the previous case, to speed-up the calculation we start at $t=0$ with a larger population of neutrinos, but still without any ELN crossing anywhere. Due to the smallness of $\Gamma$ the creation of a crossing in the decoupling zone is also much slower (notice also the longer transient). Without the presence of a  crossing, fast conversions cannot develop, as one observes until $t=800$. At later times, when a crossing is generated, fast conversions develop in the decoupling zone (see the wiggles in the snapshot at $t=1600$), producing a sudden discontinuity in the neutrino content in the free-streaming zone. Conversions are observed only for the forward modes. This is a consequence of the conservation of flavor lepton number \cite{Dasgupta:2017oko}. Indeed, the total lepton number is coming only from the backward modes, since the excess of the $\bar\nu_e$ over $\nu_e$ for the forward modes is negligible. Further, since the collisions are weaker than refractive effects, modes are not efficiently repopulated towards the equilibrium value. The oscillated forward modes then propagate towards larger $z$ (see snapshot at $t=2400$), and the effects of fast conversions can reach the free-streaming zone.

\emph{Discussion and conclusions.\,}-- Fast neutrino flavor conversions are possible near the SN core, where the angular distributions of the ELN flux, i.e., the difference of the $\nu_e$ and $\bar{\nu}_e$ fluxes, may harbor a crossing. This region is the same in which neutrinos decouple from the matter, so that they still feel residual collisions. We have studied this in a simple one-dimensional model with two momentum modes that allows us to calculate effects of neutrino flavor conversions and collisions in a consistent manner. We find that for collision rates that are significantly smaller than the neutrino potential, collisions create the conditions for fast conversions but do not damp them. Unexpectedly, state-of-art SN simulations seem to suggest that the neutrino potential indeed dominates over the collisional rate in the SN core. Drawing the insights from our model, this dominance implies that once fast conversions are generated in the decoupling zone they will propagate everywhere. With the possibility of such fast conversions, the neutrino fluxes found by SN simulations, computed without including flavor oscillations, may not be representative of reality. 

Our finding motivates a detailed analysis of current SN simulations to understand if the conditions for fast conversions are indeed generated by collisions. A dedicated analysis of angle distributions of the neutrino radiation field  for several spherically symmetric (1D) supernova simulations has not found any crossing in the ELN near the neutrinosphere~\cite{Tamborra:2017ubu}. 
However, in 3D models one expects Lepton Emission Self-sustained Asymmetry~\cite{Tamborra:2014aua} to produce a large-scale dipolar pattern in the ELN emission, which may lead to an ELN crossing (see also~\cite{OConnor:2018tuw}). This can trigger fast conversions, with possibly drastic impact on further evolution of the SN. One would need new techniques to include the effect of fast conversions into already challenging supernova simulations. This task, while obviously very challenging, may be necessary to obtain an accurate description of the supernova dynamics and neutrino fluxes.

\emph{Acknowledgments.\,}-- We acknowledge useful discussions with Hans-Thomas Janka. The work of F.C. is supported partially by the Deutsche Forschungsgemeinschaft through Grant No. EXC 153 (Excellence Cluster ``Universe'') and Grant No. SFB 1258 (Collaborative Research Center ``Neutrinos, Dark Matter, Messengers'') as well as by the European Union through Grant No. H2020-MSCA-ITN-2015/674896 (Innovative Training Network ``Elusives''). The work of B.D. is partially supported by the Dept. of Science and Technology of the Govt. of India through a Ramanujan Fellowship and by the Max-Planck-Gesellschaft through a Max-Planck-Partnergroup. The work of A.M. is supported by the Italian Istituto Nazionale di Fisica Nucleare (INFN) through the ``Theoretical Astroparticle Physics'' project and by Ministero dell'Istruzione, Universit{\`a} e Ricerca (MIUR). A.M. acknowledges support from the Alexander von Humboldt Foundation for  visits in Hamburg and Munich, where part of this work was done. The work of G.S. was supported by the Deutsche Forschungsgemeinschaft through the Collaborative Research Centre SFB 676 Particles, Strings and the Early Universe.   

\onecolumngrid

\appendix 
\section{Appendix A: Calculation of $\Gamma_{\nu}$}
\label{sec:appen}
Assuming a neutrino energy $E_\nu$ and a distance from the supernova centre $r$, the inverse of the scattering rate for the process $\nu_e+n\rightarrow e^-+p$ ($\bar{\nu}_e+p\rightarrow e^+n$)  is given by \cite{Bruenn:1985en}
\begin{equation}
\Gamma^{-1}_{\barparenb{\nu}_e}(E_\nu,r)={\frac{G^2}{\pi}n_{n\,(p)}(g_V^2+3g_A^2)(E_\nu+Q)^2\left(1-\frac{m_e^2}{(E_{\nu}+Q)^2}\right)}\,,
\label{collision_rate_nue}
\end{equation}
where $m_e$ is the mass of the electron, $G^2=5.18\times10^{-44}$ MeV$^{-2}$ cm$^2$,
$Q=1.2935$ MeV, $g_V=1$, $g_A=1.23$ and $n_{n\,(p)}(r)$ is the neutron (proton) density at a distance $r$.
In Eq.\,\ref{collision_rate_nue} we are neglecting the decrease in the scattering rate due to nucleon and electron final state blocking. Such an effect is relevant when temperatures are lower than degeneracy, e.g., when matter density becomes of the order of the nuclear density. However, since our purpose is a comparison with $\mu$ (see Fig.~1), we show that  $\mu$ dominates over $\Gamma$ even with such assumptions.

For each $r$ we calculate an average of the scattering rate using the neutrino density as weight function
\begin{equation}
\langle {\Gamma}_{\barparenb{\nu}_e}(r) \rangle =\frac{\int_0^\infty dE_\nu\Gamma_{\barparenb{\nu}_e}(E_\nu,r)n_{\barparenb{\nu}_e}}{\int_0^\infty dE_\nu n_{\barparenb{\nu}_e}}\,.
\label{collision_rate_nue_average}
\end{equation}
A similar average is performed also in the calculation of the neutrino potential $\mu(r)$. These energy-averaged quantities are shown in the main text in Fig.~\ref{fig:profiles}, where we omit the notation $\langle \cdots \rangle$ for typographic clarity.

\section{Appendix B: Numerical solution of the equation of motion}

Equation (1) in the main text represents a partial differential equation in z and t. In order to solve it we developed two different numerical codes. In the first case we used the routine D03PFF belonging to the NAG Fortran Library \cite{NAG1,NAG2}. D03PFF integrates a system of linear or nonlinear convection-diffusion equations in one space dimension, with optional source terms. The integration in time is from $t_0 = 0$ to $t_{\rm out} = 2500$, over a space interval $z \in [0,2800]$. We used 6000 mesh points in z and we adopted a $10^{-9}$ relative and absolute error tolerances for the local error test in the time integration. We checked that increasing the accuracy or the number of mesh points does not change the evolution of $\varrho(z,t)$. We provided the initial values for $\varrho(z,t)$ at $t = t_0$. For each $t$ we set the boundary conditions at $z = 0$ and $z = 2800$ to be equal to the initial conditions at the same $z$. Therefore, for the cases shown in Fig.~\ref{fig:xc01mu0}-\ref{fig:xc01} we set $\varrho_{\nu_e,\bar{\nu}_e}(0,t)=\varrho_{\nu_e,\bar{\nu}_e}(2800,t)=0$, whereas for the case reported in Fig.~\ref{fig:xc00001} we set $\varrho_{\nu_e}(0,t)=1.0$, $\varrho_{\bar{\nu}_e}(0,t)=0.8$, $\varrho(2800,t)=0$. In all cases we set $\varrho_{\nu_x,\bar{\nu}_x}(0,t)=\varrho_{\nu_x,\bar{\nu}_x}(2800,t)=0$. Note that the toy model is built so that the flavor instability never reaches the boundaries in the time interval $[t_0,t_{\rm out}]$. This choice prevents the generation of numerical artifacts. Finally, we checked that this method gives the same results when increasing the number of grid points in $z$ or the precision in the time evolution. 

The second independent code is based on Mathematica \cite{Mathematica} and has been developed to solve equation (1) for an arbitrary number of momentum modes, velocity projections on the $z$-direction, scattering rates $\Gamma_{\mathbf{p}}(z)$, equilibrium occupation number matrices $\varrho^{\rm eq}_\mathbf{p}$ and profiles for the neutrino potential $\mu(z)$. It specifically uses the predefined NDSolve routine for which typically $10^4$ grid points have been used in time and $z$-direction. In particular, results for the case shown in Fig.~\ref{fig:xc00001} were compared between the two codes and no significant differences
were found.

\vspace{0.5cm}

\bibliographystyle{JHEP}
\bibliography{biblio}

\end{document}